\documentstyle[11pt,
epsfig]{article}

\begin{document}

\relax
\renewcommand{\theequation}{\arabic{section}.\arabic{equation}}

\def\be{\begin{equation}}
\def\ee{\end{equation}}
\def\bs{\begin{subequations}}
\def\es{\end{subequations}}
\def\calm{{\cal M}}
\def\lx{\lambda}
\def\ex{\epsilon}
\def\Lx{\Lambda}

\newcommand{\tl}{\tilde t}
\newcommand{\ttt}{\tilde T}
\newcommand{\rhot}{\tilde \rho}
\newcommand{\ptt}{\tilde p}
\newcommand{\drho}{\delta \rho}
\newcommand{\drhot}{\delta {\tilde \rho}}
\newcommand{\dchi}{\delta \chi}
\newcommand{\A}{A}
\newcommand{\B}{B}
\newcommand{\mmu}{\mu}
\newcommand{\mnu}{\nu}
\newcommand{\ii}{i}
\newcommand{\jj}{j}
\newcommand{\jl}{[}
\newcommand{\jr}{]}
\newcommand{\ml}{\sharp}
\newcommand{\mr}{\sharp}

\newcommand{\da}{\dot{a}}
\newcommand{\db}{\dot{b}}
\newcommand{\dn}{\dot{n}}
\newcommand{\dda}{\ddot{a}}
\newcommand{\ddb}{\ddot{b}}
\newcommand{\ddn}{\ddot{n}}
\newcommand{\pa}{a^{\prime}}
\newcommand{\pn}{n^{\prime}}
\newcommand{\ppa}{a^{\prime \prime}}
\newcommand{\ppb}{b^{\prime \prime}}
\newcommand{\ppn}{n^{\prime \prime}}
\newcommand{\fda}{\frac{\da}{a}}
\newcommand{\fdb}{\frac{\db}{b}}
\newcommand{\fdn}{\frac{\dn}{n}}
\newcommand{\fdda}{\frac{\dda}{a}}
\newcommand{\fddb}{\frac{\ddb}{b}}
\newcommand{\fddn}{\frac{\ddn}{n}}
\newcommand{\fpa}{\frac{\pa}{a}}
\newcommand{\fpb}{\frac{\pb}{b}}
\newcommand{\fpn}{\frac{\pn}{n}}
\newcommand{\fppa}{\frac{\ppa}{a}}
\newcommand{\fppb}{\frac{\ppb}{b}}
\newcommand{\fppn}{\frac{\ppn}{n}}

\newcommand{\pt}{\tilde{p}}
\newcommand{\rhb}{\bar{\rho}}
\newcommand{\pb}{\bar{p}}
\newcommand{\pbb}{\bar{\rm p}}
\newcommand{\rht}{\tilde{\rho}}
\newcommand{\kt}{\tilde{k}}
\newcommand{\kb}{\bar{k}}
\newcommand{\wt}{\tilde{w}}

\newcommand{\dA}{\dot{A_0}}
\newcommand{\dB}{\dot{B_0}}
\newcommand{\fdA}{\frac{\dA}{A_0}}
\newcommand{\fdB}{\frac{\dB}{B_0}}

\def\be{\begin{equation}}
\def\ee{\end{equation}}
\def\bs{\begin{subequations}}
\def\es{\end{subequations}}
\newcommand{\een}{\end{subequations}}
\newcommand{\ben}{\begin{subequations}}
\newcommand{\beq}{\begin{eqalignno}}
\newcommand{\eeq}{\end{eqalignno}}

\def \lta {\mathrel{\vcenter
     {\hbox{$<$}\nointerlineskip\hbox{$\sim$}}}}
\def \gta {\mathrel{\vcenter
     {\hbox{$>$}\nointerlineskip\hbox{$\sim$}}}}

\def\g{\gamma}
\def\mpl{M_{\rm Pl}}
\def\ms{M_{\rm s}}
\def\ls{l_{\rm s}}
\def\l{\lambda}
\def\m{\mu}
\def\n{\nu}
\def\a{\alpha}
\def\b{\beta}
\def\gs{g_{\rm s}}
\def\d{\partial}
\def\co{{\cal O}}
\def\sp{\;\;\;,\;\;\;}
\def\r{\rho}
\def\dr{\dot r}

\def\e{\epsilon}
\newcommand{\NPB}[3]{\emph{ Nucl.~Phys.} \textbf{B#1} (#2) #3}   
\newcommand{\PLB}[3]{\emph{ Phys.~Lett.} \textbf{B#1} (#2) #3}   
\newcommand{\ttbs}{\char'134}        
\newcommand\fverb{\setbox\pippobox=\hbox\bgroup\verb}
\newcommand\fverbdo{\egroup\medskip\noindent%
                        \fbox{\unhbox\pippobox}\ }
\newcommand\fverbit{\egroup\item[\fbox{\unhbox\pippobox}]}
\newbox\pippobox
\def\tr{\tilde\rho}
\def\lb{w}
\def\bbox{\nabla^2}
\def\mt{{\tilde m}}
\def\rct{{\tilde r}_c}

\def \lta {\mathrel{\vcenter
     {\hbox{$<$}\nointerlineskip\hbox{$\sim$}}}}
\def \gta {\mathrel{\vcenter
     {\hbox{$>$}\nointerlineskip\hbox{$\sim$}}}}

\noindent
\begin{flushright}

\end{flushright} 
\vspace{1cm}
\begin{center}
{ \Large \bf Mirage effects on the brane\\} 
\vspace{0.5cm}
{P. S. Apostolopoulos, N. Brouzakis, E. N. Saridakis and N. Tetradis} 
\\
\vspace{0.5cm}
{\it University of Athens, Department of Physics, Nuclear and Particle Physics Section,\\ 
Panepistimiopolis, Zografos 15771, Athens, Greece.} 
\\
\vspace{1cm}
\abstract{We discuss features of the brane cosmological evolution that arise
through the presence of matter in the bulk. As these deviations from the
conventional evolution are not associated with some observable 
matter component on the brane, we characterize them as mirage effects.
We review an example of expansion that can be attributed to
mirage non-relativistic matter (mirage cold dark matter) on the brane.
The real source of the evolution is an anisotropic bulk fluid with negative
pressure along the extra dimension. 
We also study the general problem of exchange of 
real non-relativistic matter between the brane and the
bulk, and discuss the related mirage effects.
Finally, we derive the brane cosmological evolution within a bulk that contains
a global monopole (hedgehog) configuration. This background induces a
mirage curvature term in the effective Friedmann equation, 
which can cause a brane 
Universe with positive spatial curvature to expand forever. 
\\
\vspace{1cm}
PACS number: 98.80.-k, 98.80.Cq, 11.25.-w, 11.27.+d
} 
\end{center}

\newpage

\section{Introduction}
\setcounter{equation}{0}

In the context of the Randall-Sundrum model \cite{rs},
the Universe is identified with a four-dimensional hypersurface
(a 3-brane) in a five-dimensional bulk with negative cosmological
constant (AdS space). The geometry is non-trivial (warped)
along the fourth spatial dimension, so that an effective localization of
low-energy gravity takes place near the brane.
For low matter densities on the brane and a pure AdS bulk, 
the cosmological evolution as 
seen by a brane observer reduces to the standard
Friedmann-Robertson-Walker cosmology 
\cite{binetruy,csaki,kraus}.
(For recent reviews,
with extensive lists of references, see ref. \cite{brax}.)
The only novel feature is a contribution to the effective Friedmann equation
that has the form of a conserved radiation term. 
This is induced by the gravitational field in the bulk space and
can be characterized as a mirage effect, not associated with a matter
component on the brane \cite{binetruy,mirage,hebecker}. 

In all five-dimensional effective actions derived from 
string theory there is a variety of fields in the bulk. 
Typical effective theories are gauged versions of five-dimensional
supergravities coupled to 
four-dimensional boundary theories with gauge and 
matter fields. In addition, there could be various moduli fields with 
potential terms in the bulk as well as on the boundary.
(For a partial list of related works see ref. \cite{supergravity}.) 
It seems, therefore, that the Randall-Sundrum model is a simplification 
of the general case. Especially in the context of cosmology,
it is natural to expect that the bulk degrees of freedom will be excited
by the available energy density. The simplest example is provided by
the transfer of energy into the bulk through the decay of thermalized
brane particles to
bulk gravitons. The cosmological evolution of ref. \cite{binetruy}, in which
all the energy density is localized on the brane, while the bulk energy
momentum tensor includes only a negative cosmological constant, is 
a simplification of the much more complicated realistic scenario.

If the bulk contains some matter component in addition to the negative
cosmological constant the cosmological evolution is modified.\footnote{
Modifications of the evolution can also appear because of the 
inclusion of additional interactions in the gravitational sector,
such as an induced gravity term on the brane \cite{induced}, 
or a Gauss-Bonnet term in the bulk \cite{gauss}. We consider only a standard
Einstein term in the bulk, and concentrate on non-trivial matter components.} 
For example, it is possible to have energy exchange between the brane and the bulk 
\cite{vandebruck,tet,tomaras}. 
Also the presence of a fluid in the bulk can alter the expansion on the brane. 
The
modifications can be attributed to mirage matter components 
on the brane \cite{bulk1,bulk2,bulk3}.

In the case of an empty bulk,
the brane evolution can be discussed either in a coordinate system
(system A) in which the brane is located at a fixed value of the fourth spatial
coordinate and the bulk is time-dependent \cite{binetruy,csaki}, or
in a different coordinate system (system B) 
in which the bulk is static and the brane
is moving \cite{kraus}. In the latter case, the bulk metric is 
five-dimensional Schwarzschild-AdS \cite{birm}. The two points of view are
equivalent \cite{gregory}. 

The interpolation between the two coordinate 
systems can be employed in order to construct examples of brane evolution
in a bulk that contains a fluid in addition to the negative cosmological
constant. A configuration that can be
characterized as star-AdS was considered in ref. \cite{bulk1}:
The bulk fluid is spherically symmetric, denser at the origin, while its
density goes to zero at a finite value of the radial coordinate.
Another well studied case assumes the presence of a 
radiation field in the bulk, with the resulting bulk metric having the
Vaidya-AdS form \cite{wang}. This background permits the study of 
radiation or graviton emission by the brane towards the
bulk \cite{hebecker,vaidya}.
One can also consider the possibility of radiation or graviton absorption 
by the
brane \cite{bulk2}. Both the above examples involve mirage contributions in
the effective Friedmann equation.

In this work we study other mirage effects that could have an interesting
physical interpretation. 
We do not focus on studying a specific supergravity model. We are more 
interested in exploring novel cosmological behaviour, starting from
a simple ansatz that can lead to a complete solution of the (often
formidable) Einstein equations. 

In the following section, we briefly summarize known results for
the mirage radiation, employing the generalized
Vaidya-AdS metric in the bulk.
We also discuss in some detail the observation of ref. \cite{bulk2} 
that a certain form of the generalized Vaidya-AdS metric
results in a mirage term characteristic of 
non-relativistic matter (mirage cold dark matter). 
We show that the 
real source of the evolution is an anisotropic bulk fluid with negative
pressure along the extra dimension. Its energy-momentum tensor resembles
that of a global monopole configuration in four dimensions. 
In section 3 we study the related problem of the exchange of real 
non-relativistic matter between the brane and the bulk.
We discuss the mirage effects arising in this system.
In section 4
we study the brane evolution induced by a global monopole background 
in the five-dimensional
bulk. We find that a mirage
curvature term appears on the brane.

\section{Mirage radiation and cold dark matter}

We start by reviewing some known results on mirage radiation.
We consider a bulk metric of the form
\begin{equation}
ds^{2}=-n^{2}(u,r)\, du^{2}+2\ex\,du \,dr+ r^{2}d\Omega_k^2,
\label{metric2}
\end{equation}
where 
\be
n^2(u,r)=\frac{1}{12M^3}\Lx r^2+k-\frac{1}{6\pi^2M^3} \frac{\calm(u,r)}{r^2} 
\label{ns} \ee
and $d\Omega^2_k$ is the metric of a
maximally symmetric three-dimensional space ($k=-1,0,1$).
This is a generalized Vaidya-AdS metric \cite{wang}.
The cosmological constant is equal to $-\Lx$, while $M$ is the fundamental 
Planck constant. 
The parameter $\ex$ takes the
values $\ex=\pm 1$.
In studies of graviton emission from the brane it is usually assumed
that $\calm=\calm(u)$ and $\ex=1$. 
Our discussion is more general, as it allows for 
an additional dependence of $\calm$ on $r$. It can also account for
energy absorption by the brane when $\ex=-1$.

The energy-momentum tensor
that satisfies the Einstein equations is
\begin{eqnarray}
{T}^{0}_{~0} = {T}^{4}_{~4} &=& \Lx -\frac{1}{2\pi^2} \frac{\calm_{,r}}{r^3}
\label{t002} \\
{T}^{1}_{~1} = {T}^{2}_{~2} = {T}^{3}_{~3} 
&=& \Lx -\frac{1}{6\pi^2} \frac{\calm_{,rr}}{r^2}
\label{t112} \\
{T}^{0}_{~4} &=& \frac{1}{2\pi^2} \frac{\calm_{,u}}{r^3},
\label{t042} 
\end{eqnarray} 
where the subscripts indicate derivatives with respect to
$r$ and $u$. The matter component that is added to the cosmological
constant satisfies the
various energy conditions 
if 
$\ex \calm_{,u} \geq 0$, $\calm_{,r}\geq 0$,
$\calm_{,rr}\leq 0$, $\calm_{,r}\geq -r\calm_{,rr}/3$ \cite{wang}.

For the discussion of the cosmological evolution on the brane 
we consider a coordinate system 
in which the metric takes the form
\begin{equation}
ds^2=\gamma_{ab}dx^a dx^b+d\eta^2 
=-m^2(\tau,\eta )d\tau^2
+a^2(\tau,\eta )d\Omega_k ^2+d\eta ^2.
\label{sx2.1ex2}
\end{equation}
The brane is located at $\eta=0$, while we
identify the half-space $\eta>0$ with the half-space
$\eta<0$. We also redefine the time, so as to 
set $m(\tau,\eta=0)=1$.
Through an appropriate coordinate transformation 
\begin{equation}
u=u({\tau},\eta ),\qquad r=a({\tau},\eta )  \label{sx2.42}
\end{equation}
the metric (\ref{metric2})
can be written in the form of eq. (\ref{sx2.1ex2}).

The equations governing the cosmological evolution on the brane 
are \cite{vaidya,bulk2}
\begin{eqnarray}
H^2
&=&\left( \frac{\dot{R}}{R} \right)^2=
\frac{1}{144 M^6}\rht^2+ \frac{1}{6\mpl^2}\rht
+\frac{1}{6\pi^2M^3}\frac{\calm(\tau,R)}{R^4}
-\frac{k}{R^2}
+\lx
\label{hubble22} \\
\dot{\rht}+3 H (\rht+\pt)&=&
-\frac{12M^3}{\pi^2V}\frac{\dot{\calm}(\tau,R)}{R^4}
\frac{1}{1-\ex\frac{12M^3H}{V}+\frac{\rht}{V}}, 
\label{energ2}
\end{eqnarray} 
where $\rht$, $\pt$ are the brane energy density and pressure
and $R(\tau)=a(\tau,\eta=0)$. 
We have also denoted $\calm(\tau,R)\equiv \calm(u(\tau,\eta=0),r=R)$.
The dot denotes a partial derivative with respect to
$\tau$. The above equations have the general form expected for 
brane cosmologies \cite{general}.
The requirement $\ex \calm_{,u} \geq 0$, imposed by the energy conditions,
indicates that we must associate $\ex=1$ with energy outflow (for which
$\dot{\calm}\equiv\partial \calm(\tau,R)/\partial \tau>0$) and $\ex=-1$ 
with energy inflow (for which $\dot{\calm}<0$).
We have also defined
$\mpl^2=12M^6/V$, where $V$ is the brane tension. The bulk cosmological 
constant and the brane tension are fine tuned so that the effective
cosmological constant vanishes: 
$\lx=(V^2/12M^3-\Lx)/12M^3=0$.
In the low-energy regime the last factor in the r.h.s. of eq. (\ref{energ2})
becomes 1, so that the same equations describe energy outflow or inflow,
depending on the sign of $\dot{\calm}$.

For $\calm=\calm(\tau)$ and large $R$
we can put eqs. (\ref{hubble22}), (\ref{energ2}) in the form
\begin{eqnarray}
H^2 &=&\left( \frac{\dot{R}}{R} \right)^2=
\frac{1}{6\mpl^2}(\rht+\rht_r)
-\frac{k}{R^2}
\label{hubble223} \\
\dot{\rht}+3 H (\rht+\pt)&=&-(\dot{\rht}_r+4 H \rht_r),
\label{energ23}
\end{eqnarray}
with 
$\rht_r=12M^3\calm(\tau)/(\pi^2 V R^4)$.
These equations describe an expanding Universe in which 
brane matter can be transformed to mirage radiation, or the opposite,
while the total energy is conserved. 
In the high-energy regime the two possible values of $\ex$ result in 
different forms of eq. (\ref{energ2}).

In the case $\calm=\calm(r)$ the brane matter and the mirage component
evolve independently. The metric of eqs. (\ref{metric2}), (\ref{ns})
can be written in the form 
\begin{equation}
ds^{2}=-n^{2}(r)\, dt^{2} +n^{-2}(r)\, dr^{2} + r^{2}d\Omega_k^2,
\label{metric3}
\end{equation}
where 
\be
n^2(r)=\frac{1}{12M^3}\Lx r^2+k-\frac{1}{6\pi^2M^3} \frac{\calm(r)}{r^2} 
\label{ns2} \ee
and $\partial u/\partial t=1$, $\partial u/\partial r= \ex/n^2(r)$.
The non-zero components of the
energy-momentum tensor that satisfies the Einstein equations for this
metric are given by 
\begin{eqnarray}
{T}^{0}_{~0} = {T}^{4}_{~4} &=& \Lx -\frac{1}{2\pi^2} \frac{\calm_{,r}}{r^3}
\label{t002a} \\
{T}^{1}_{~1} = {T}^{2}_{~2} = {T}^{3}_{~3} 
&=& \Lx -\frac{1}{6\pi^2} \frac{\calm_{,rr}}{r^2}.
\label{t112a}
\end{eqnarray} 
For $\calm$=constant we recover the 
standard Schwarzschild-AdS metric, as expected. The mirage component is 
identified with the mirage radiation that evolves independently from the 
brane matter.

As was pointed out in ref. \cite{bulk2}, 
the generalized Vaidya metric of eqs. (\ref{metric2}), (\ref{ns})
allows for a non-trivial dependence of $\calm$ on $R$.
We can assume that 
$\calm=\zeta \, \gamma(\tau) R^n$ with $\zeta=\pm 1$, $n$ integer, and
$\gamma(\tau)$ a positive-definite function.
The various energy conditions constrain the possible values of 
$n$ and $\zeta$. Apart from the case $(n,\zeta)=(0,1)$ that we discussed
above, we have the possibilites $(n,\zeta)=(1,1),(-1,-1),(-2,-1)$.
The last two cases result in mirage terms in the effective
Friedmann equation that fall off faster than $R^{-4}$. For this reason
they are insignificant in the low-energy regime for large $R$. 
We concentrate on the case $\calm=\gamma(\tau) R$. 

In the low-energy regime, the evolution equations (\ref{hubble22}), 
(\ref{energ2}) can be written as
\begin{eqnarray}
H^2 &=&\left( \frac{\dot{R}}{R} \right)^2=
\frac{1}{6\mpl^2}(\rht+\rht_m)
-\frac{k}{R^2}
\label{hubble224} \\
\dot{\rht}+3 H (\rht+\pt)&=&-(\dot{\rht}_m+3 H \rht_m),
\label{energ24}
\end{eqnarray}
with 
$\rht_m=12M^3\gamma(\tau)/(\pi^2 V R^3)$.
These equations describe the transformation of brane matter to mirage 
dust, or
the opposite, in an expanding Universe. 
The bulk fluid that is associated with the necessary energy-momentum
tensor for such a solution
does not have an obvious physical interpretation.
In order to obtain some intuition, we
assume that $\gamma$ is time independent, so that there is
no energy exchange between the brane and the bulk.
In the coordinate system $(t,r)$, in which the metric has the 
form of eqs. (\ref{metric3}), (\ref{ns2}), 
the bulk contains, apart from the negative cosmological constant, 
an anisotropic fluid 
with (see eq. (\ref{t002a}), (\ref{t112a}))
\begin{eqnarray}
T^0_{~0}=T^4_{~4}&=&-\frac{\gamma}{2\pi^2r^3}
\nonumber \\
T^1_{~1}=T^2_{~2}=
T^3_{~3}&=&0.
\label{last} \end{eqnarray}
This is equivalent to $\calm(r)=\gamma r$.
If we parametrize the bulk energy-momentum tensor as 
$T^A_{~B}={\rm diag} (\rho,p,p,p,{\rm p})$, 
the fluid must have an equation of state
$p=0$, ${\rm p}= -\rho$.

In four dimensions, there is a physical system 
with an equation of state similar to 
the one above. At large distances from the center of
a global monopole, the energy-momentum
tensor is given by
$T^t_{~t}\simeq T^r_{~r}\sim - r^{-2}$ and $T^\theta_{~\theta}
=T^\phi_{~\phi}\simeq 0$ \cite{vilenkin}. It is possible that an analogous
configuration in five dimensions may result in the required form of 
the equation of state. We explore this possibility in
section 4.

A related question in brane cosmology concerns the
possibility that real 
non-relativistic matter is exchanged between the bulk and the brane. 
In this case the total amount of cold dark matter is not conserved 
on the brane, as particles can either escape to the bulk or be absorbed
by the brane. 

This question is related to an interesting problem in 
brane cosmology. It has been argued that the absoprtion of energy by
the brane may lead to periods of accelerated expansion \cite{tet,tomaras}.
An important issue is whether the absorbed matter may 
have non-negative pressure, while the brane expansion remains accelerated. 
This could lead to the elimination of the inflaton field as a necessary
ingredient of inflation. 
It has been shown, however, on general grounds that the 
bulk matter must have negative pressure for any acceleration to occur
\cite{general}. We are interested in addressing this issue through
an explicit calculation in a specific model.

In the absence of a cosmological constant 
a pressureless bulk gas can be discussed in terms of 
the Tolman-Bondi metric \cite{tb}.
This problem is the non-relativistic analogue 
of the relativistic case studied through the Vaidya-AdS bulk metric 
(\ref{metric2}).
The cosmological evolution and the induced mirage effects on the brane 
are studied in the following
section. Particular emphasis is put on the possible link 
between acceleration and inflow of non-relativistic matter.

\section{Real cold dark matter}

A non-relativistic perfect
fluid on the brane is described by the equation of state $\pt=0$.
In the bulk the appropriate metric for the description of a
pressureless, inhomogeneous fluid is the  
Tolman-Bondi metric \cite{tb}. We employ here a generalization 
that allows for 
a non-zero negative cosmological constant. 

The bulk metric can be written in the form 
\begin{equation}
ds^{2}=-dt^2+b^2(t,r)dr^2+S^2(t,r)d\Omega_k^2,
\label{metrictb}
\end{equation}
where $d\Omega^2_k$ is the metric of a
maximally symmetric three-dimensional space ($k=-1,0,1$).
The function $b(r,t)$ is given by
\begin{equation}
b^2(t,r)=\frac{S^2_{,r}(t,r)}{k+f(r)},
\label{brttb} \end{equation}
where the subscript denotes differentiation with respect to $r$, and
$f(r)$ is an arbitrary function.
The bulk energy momentum tensor has the form
\begin{equation}
T^A_{~B}={\rm diag} \left(\Lx-\rho(t,r),\, \Lx,\, \Lx,\, \Lx,\, \Lx  \right).
\label{enmomtb} \end{equation}
The bulk fluid consists of successive shells marked by $r$, whose
local density $\rho$ is time-dependent. 
The function $S(t,r)$ describes the location of the shell marked by $r$
at the time $t$. 
The Einstein equations reduce to 
\begin{eqnarray}
S^2_{,t}(t,r)&=&\frac{1}{6\pi^2M^3}\frac{\calm (r)}{S^2}-\frac{1}{12M^3}\Lx S^2
+f(r)
\label{tb1} \\
\calm_{,r}(r)&=&2\pi^2 S^3 \rho \, S_{,r}.
\label{tb2} \end{eqnarray}
The generalized mass function $\calm(r)$ of the bulk fluid incorporates the
contributions of all shells between 0 and $r$. It can be obtained through
the integration of eq. (\ref{tb2}). Because of energy conservation
it is independent of $t$, while $\rho$ and $S$ depend on both $t$ and $r$.

It is obvious that the solution we are considering
is appropriate for describing a matter distribution only if the
r.h.s. of eq. (\ref{tb1}) remains positive. The functions $\calm (r)$, $f(r)$
must have a form that guarantees this at $t=0$. The subsequent 
evolution for an expanding fluid ($S_{,t}>0$) inevitably leads to 
$S_{,t}$ becoming zero for a certain value of $r$ at some time $t$. 
This could happen either
for $r\to \infty$ if the initial matter distribution extends over the whole
extra dimension, or at the point where the matter density becomes zero. 
(In the latter case, the metric assumes the standard
Schwarzschild-AdS form for larger $r$ \cite{birm}.) At later times the
outer regions of the matter distribution start collapsing, while the
inner ones continue their expansion. Shell crossing (characterized by
$S_{,r} < 0$) is unavoidable for a pressureless fluid. This behaviour is
caused by the form of the geodesics in an AdS space, which always lead to 
$r=0$. Non-relativistic matter collapses to this point at late times, even if
initially it has an outgoing velocity.

In our discussion
we shall not explore the full dynamics of the Tolman-Bondi-AdS metric. 
We shall consider the motion of a brane within an expanding fluid, by
choosing appropriate initial conditions and limiting the
time interval during which we follow the evolution. A simple late-time
scenario would have the brane moving outside the region of non-zero 
matter density. Then the bulk metric would take the
Schwarzschild-AdS form, 
depending only on the constant total integrated mass of the bulk fluid.

For the discussion of the cosmological evolution on the brane we proceed
similarly to section 2. We consider a coordinate system in which the metric
takes the form of eq. (\ref{sx2.1ex2}). Through an appropriate coordinate
transformation 
\begin{equation}
t=t({\tau},\eta ),\qquad r=r({\tau},\eta )  \label{sx2.42ex}
\end{equation}
the metric of eq. (\ref{metrictb})
can be written in the form of eq. (\ref{sx2.1ex2}).
Clearly $a(\tau,\eta)=S(t,r)$.
We define $R(\tau)=a(\tau,\eta=0)$. This quantity corresponds to the
scale factor of the brane. In the same time, through
the relation $a(\tau,\eta=0)=S(t(\tau,\eta=0),r(\tau,\eta=0))$ it can be
interpreted as the location of the brane in terms of the coordinate $S$.
Through the relation $t=t(\tau,\eta)$ a connection can be established between
the time coordinates $t$ and $\tau$ at the location of the brane $\eta=0$.
Then the quantity $r(\tau,\eta=0)$ denotes the shell whose location 
coincides with that of the brane. Clearly $R(\tau)$ and 
$r(\tau,\eta=0)$ do not have the same $\tau$-dependence in general.
Depending on the values of $\dot{R}(\tau)$ and $\dot{r}(\tau,\eta=0)$,
the bulk gas can move faster or slower than the brane. In the first case,
bulk matter has to be absorbed by the brane (as the brane essentially
forms the boundary of the AdS space), while in the second energy must
be emitted into the bulk by the brane. We shall see an explicit example
of this behaviour in the following.

At the location of the brane (where $S=R$) we find 
\begin{eqnarray}
t_{,\tau}&=&
\frac{1}{W}
\left(-\delta \dot{R} \sqrt{-W+k+f} 
+\epsilon \sqrt{k+f}\sqrt{\dot{R}^2+W} \right)
\label{dtdtau} \\
t_{,\eta}&=&
\frac{1}{W}
\left(-\dot{R} \sqrt{k+f} 
+\delta \epsilon \sqrt{-W+k+f}\sqrt{\dot{R}^2+W} \right),
\label{dtdeta} 
\end{eqnarray}
with
\begin{equation}
W=-\frac{1}{6\pi^2M^3}\frac{\calm }{S^2}+\frac{1}{12M^3}\Lx S^2+k
\label{ww} \end{equation}
and the dot denoting a derivative with respect to $\tau$.
The parameters $\delta$, $\epsilon$ take the values $\pm 1$.
For $S_{,t}>0$ $(<0)$ we must choose $\delta=1$ $(-1)$.
The value of $\epsilon$ is fixed by the requirement $t_{,\tau}>0$, so
that time flows in the same direction both for brane and bulk observers.  
For an expanding bulk fluid ($\delta =1$) we must take $\epsilon=1$.

The bulk
energy-momentum tensor as measured by a brane observer (in the frame
of eq. (\ref{sx2.1ex2})) has
\begin{eqnarray}
T^0_{~0}&=&\Lx-\rho\,\, t_{,\tau}^2
\label{emtbr1} \\
T^0_{~4}&=&-\rho\,\, t_{,\tau}\,\, t_{,\eta} 
\label{emtbr2} \\
T^4_{~4}&=&\Lx+\rho \,\, t_{,\eta}^2
\label{emtbr3} \\
T^1_{~1}&=&T^2_{~2}=T^3_{~3}=\Lx.
\label{emtbr4} \end{eqnarray}

The equations governing the cosmological evolution on the brane
can be obtained in complete analogy to refs. \cite{bulk1,bulk2}.
It can be shown that, at the location of the brane, they take the form
\begin{eqnarray}
H^2
&=&\left( \frac{\dot{R}}{R} \right)^2=
\frac{1}{144 M^6}\rht^2+ \frac{1}{6\mpl^2}\rht
+\frac{1}{6\pi^2M^3}\frac{\calm(r(\tau,\eta=0))}{R^4}
-\frac{k}{R^2}
\label{hubble225} \\
\dot{\rht}+3 H \rht&=&-2T^0_{~4}(t(\tau,\eta=0),r(\tau,\eta=0)).
\label{energ25}
\end{eqnarray} 
(We have set the effective cosmological constant to zero.)

The cosmological evolution described by the equations in this section 
depends on the value of the curvature parameter $k$. We expect that
this dependence does not display novel qualitative features, 
other than those encountered in conventional cosmology. 
In particular, eqs. (\ref{tb1}), (\ref{tb2}) do not depend on
$k$. In eqs. (\ref{dtdtau}), (\ref{dtdeta}), the $k$-dependence
cancels in the factors $-W+k$ and $\dot{R}^2+k$. The only explicit
dependence appears in the factor $k+f(r)$, which must be assumed to
be positive for $r$ to be space-like, according to eq. (\ref{brttb}).  
The function $f(r)$ can be linked to the initial velocity of the 
bulk fluid through eq. (\ref{tb1}).
If the fluid has a large initial velocity, the contribution from $k$ is
negligible. Otherwise, its value may induce only a quantitative modification of
the cosmological evolution. 
The only significant qualitative role played by $k$ is in
eq. (\ref{hubble225}). It determines the late time behaviour of the
cosmological evolution, similarly to conventional cosmology.

The rate of energy exchange between the brane and the bulk is determined by
the element $T^0_{~4}$ of the bulk energy-momentum tensor at the
location of the brane, as given
by eqs. (\ref{emtbr2}) and (\ref{dtdtau}), (\ref{dtdeta}), (\ref{ww}). 
The evolution of $T^0_{~4}$ with time has a complicated 
dependence on the arbitrary functions $f(r)$ and $\calm(r)$. 
This is different than in the Vaidya-AdS case discussed in the previous
section, in which the energy exchange is directly determined by
eq. (\ref{t042}) through the
assumed form of $\calm(r,u)$.\footnote{However, even in this
case there is an ambiguity: An assumed form of 
$T^0_{~4}(\tau)$ can be matched by various forms of 
$\calm(r(\tau,\eta=0),u(\tau,\eta=0))$.}
For this reason, it is technically very difficult to use as input the form of 
$T^0_{~4}(\tau)$ and deduce the required form of 
$f(r)$ and $\calm(r)$. It seems reasonable, even though not guaranteed,
that the freedom provided by the 
choice of two functions allows for an arbitrary 
form of $T^0_{~4}(\tau)$. In physical terms, 
the form of the energy exchange between
the brane and the bulk is directly related to 
the distribution and velocity of the bulk matter.

Instead of deriving the general solution of the bulk-brane system, which
seems formidable even through a numerical approach, we shall
simplify the equations through some additional assumptions, in order to 
address the connection between energy inflow and acceleration that we
discussed at the end of the last section.
At the initial time $t=0$ we can choose the coordinates
$r$ and $S$ to coincide
($S(0,r)=r$). 
We make the simplifying assumption that the bulk fluid is homogeneous.
This implies that 
$\calm(r)=\pi^2r^4 \rho_0/2$, 
with $\rho_0=\rho(t=0)$. 
The Einstein equation (\ref{tb1}) can be integrated if
$f(r)$ is known. At the time $t=0$, this function can
be linked to the initial velocity of the 
bulk fluid. The homogeneity is preserved if 
we make the additional assumption $f(r)=v_0^2\,r^2$. 
This permits as to
write eq. (\ref{tb1}) as 
\begin{equation}
\frac{s^2_{,t}}{s^2}
=\frac{1}{12M^3}\frac{\rho_0}{s^4}-\frac{1}{12M^3}\Lx  +\frac{v_0^2}{s^2},
\label{dadt} \end{equation}
with $S(t,r)=s(t)\, r$ and $s(0)=1$.
The above equation has a form very similar to the 
standard Friedmann equation, with the curvature parameter
$k$ replaced by the ``initial squared velocity'' $v^2_0$. It
can be integrated easily for given $\rho_0$, $\Lx$, $v_0$.
The bulk fluid remains homogeneous with a density 
$\rho(t)=\rho_0/s^4(t)$.

\begin{figure}[t]
 \centerline{\epsfig{figure=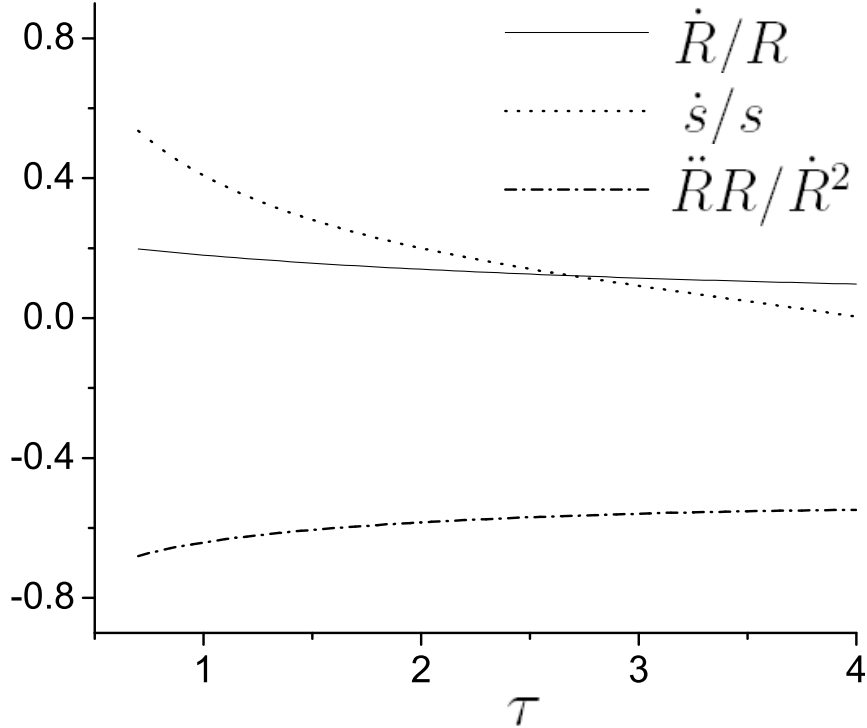,width=13cm,angle=0}}
 \caption{\it
The expansion rate $\dot{R}/R$ and the acceleration parameter
$\ddot{R}R/\dot{R}^2$ of the brane, and the expansion rate $\dot{s}/s$
of the bulk fluid for $M=1$, $\Lambda=1$, $V=\sqrt{12}$, $\rho_0=0.2$, 
$v_0=1$, $R(0)=0.5$, $s(0)=1$ and $\rht(0)=1$.
}
 \label{fig1}
 \end{figure}

In order to obtain an example of brane evolution through numerical
integration, we proceed as follows: We describe the evolution in terms
of the proper time $\tau$ measured by an observer comoving with the brane.
We concentrate on a flat brane Universe with $k=0$. From 
eqs. (\ref{dtdtau}), (\ref{dadt}) we obtain an expression
for $\dot{s}(\tau)=[s_{,t}
\,t_{,\tau}](\tau,\eta=0)$. Eqs. (\ref{hubble225}), (\ref{energ25})
determine the evolution of $R(\tau)$, $\rht(\tau)$. The value of
$T^0_{~4}$ at the location of the brane, as measured by the brane observer, 
is given by eqs. (\ref{emtbr2}), (\ref{dtdtau}), (\ref{dtdeta}).
The ratio $R(\tau)/s(\tau)=r(\tau,\eta=0)$ determines the
shell whose location coincides with that of the brane at time $\tau$.
The mass function is given by the expression $\calm(\tau)=\pi^2
r^4(\tau,\eta=0) \rho_0/2$. A solution is uniquely determined by a
choice of the parameters $M$, $\Lambda$, $V$, $\rho_0$, $v_0$ and
the initial conditions $R(0)$, $s(0)$ and $\rht(0)$.

\begin{figure}[t]
 \centerline{\epsfig{figure=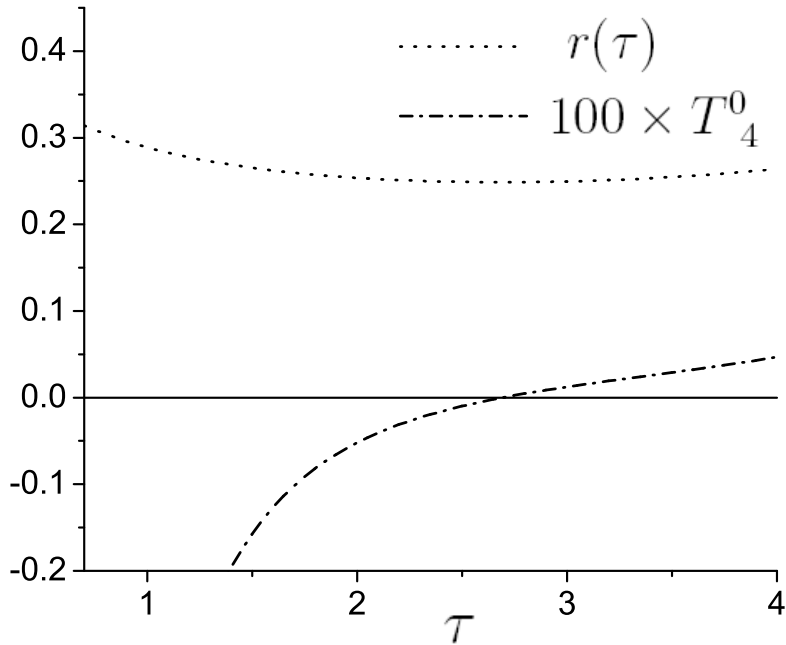,width=13cm,angle=0}}
 \caption{\it
The coordinate $r(\tau)$ of the bulk fluid shell that coincides with
the brane, and the energy flow $T^0_{~4}$ at the location of the brane. 
}
 \label{fig2}
 \end{figure}

In figs. 1 and 2 we depict some of the characteristics of the evolution 
for the choice $M=1$, $\Lambda=1$, $V=\sqrt{12}$, $\rho_0=0.2$, $v_0=1$, 
$R(0)=0.5$, $s(0)=1$ and $\rht(0)=1$. In fig. 1 we plot the expansion rate
of the brane $\dot{R}/R$ as measured by a brane observer. The same quantity
is related to the velocity of the brane as seen by a bulk observer
in the system of coordinates (\ref{metrictb}). The expansion of the bulk
fluid is given by $\dot{s}/s$. According to our previous discussion,
we have the relation $\dot{R}/R=\dot{s}/s +\dot{r}/r$, where $r(\tau)$
denotes the shell that coincides with the brane at the time $\tau$.
Fig. 1 indicates that before $\tau_{e} \simeq 2.7$ the bulk fluid expands
faster than the brane, while at later times it is overtaken by the brane.
At a time $\tau \simeq 4$ the bulk fluid stops expanding ($\dot{s}=0$).
At later times it is expected to reverse its motion, so that
shell crossing takes place.

In fig. 2 we plot the function $r(\tau)$. This has a minimumn 
at the time $\tau_{e}\simeq 2.7$, when the expansion of the bulk fluid
is overtaken by the brane motion. For $\tau < \tau_{e}$ there is 
energy flowing onto the brane from the bulk. This is expected,
as the brane essentially forms the boundary of the bulk space. When
the bulk fluid expands faster than the brane, the excess energy 
near the boundary is accumulated on it. For $\tau > \tau_{e}$ the process
is reversed. This behaviour is summarized by the time dependence of
the 04-component of the bulk energy-momentum tensor at the location of
the brane, as measured by the brane observer (eq. (\ref{emtbr2})). 

In fig. 1 we also plot the acceleration parameter 
$\ddot{R}R/\dot{R}^2$ for the brane expansion. We observe that is remains
negative for the whole evolution, irrespectively of the direction of the
energy flow. The reason can be understood in the general framework of
ref. \cite{general}. The effect of the bulk matter on the brane
evolution can be described in terms of a mirage brane component with 
effective density and 
pressure
\begin{eqnarray}
\rho_{eff}&=&\frac{12M^3}{\pi^2V}\, \frac{\calm(R)}{R^4}
\label{rhoeff} \\
p_{eff}&=&\frac{1}{3}\rho_{eff}+ \frac{8M^3}{V} \bar{p},
\label{peff} \end{eqnarray}
where $\bar{p}$ is the pressure of the bulk fluid along the extra
dimension, as measured by the brane observer. In our case this
is given by the second contribution to $T^4_{~4}$ of eq. (\ref{emtbr3}),
and is always positive. As a result, both fluids that affect the
brane evolution (the real non-relativistic matter and the mirage component)
have positive energy density and non-negative pressure. The resulting 
expansion must be decelerating. 

The example that we discussed demonstrates how the presence of a 
pressureless fluid in the bulk and the brane affects the evolution.
The brane component contributes directly 
to the effective Friedmann equation for
the expansion on the brane. The bulk component generates a mirage term,
which in general falls faster than $R^{-4}$. (This is apparent from the 
form (\ref{peff}) of the effective pressure.) 
Moreover, there is energy exchange between the brane and mirage components.
The energy exchange is connected to the relative magnitude of the expansion
rate on the brane (which can also be viewed as the motion of the brane within
the bulk) and the expansion rate of the bulk fluid. In general, the bulk
fluid in a Tolman-Bondi-AdS geometry
depends on two arbitrary functions: $\calm(r)$ and $f(r)$. 
Their form is related to the rate of energy exchange between the bulk and
the brane, as quantified by the value of $T^0_{~4}$ at the location of the
brane.\footnote{For an attempt to construct a physical
mechanism underlying the absorption of bulk massive particles by the brane, 
see ref. \cite{brito}.} 
The relation is not as explicit as in the case of 
the Vaidya-AdS metric for a radiation fluid 
(see eq. (\ref{t042})).
However, the physical picture is the same: The energy exchange between
the brane and the bulk is related to the distribution and velocity of the 
bulk matter. 

\section{Mirage curvature}
We return now to the issue of a mirage component that could behave as
non-relativistic matter (mirage cold dark matter).
In section 2 we saw that the global monopole (hedgehog) configuration has
an energy-momentum tensor of the form that could generate such a mirage
effect. In this section we explore this possibility by constructing
explicitly a global monopole configuration in a
five-dimensional AdS background,
and embedding a brane in it. 

In order to construct a global monopole (hedgehog) configuration in
five dimensions we consider a four-component field 
$\phi^\alpha$, $\alpha=1,2,3,4,$ with an $O(4)$ symmetry.
Its Lagrangian is given by
\begin{equation}
{\cal L}(\phi)=-\frac{1}{2}\phi^\alpha_{~;C}\phi_{\alpha;D}g^{CD}
-U_B(\phi)=
-\frac{1}{2}\phi^\alpha_{~;C}\phi_{\alpha;D}g^{CD}
-\frac{\lx}{4}\left(\phi^\alpha \phi_\alpha -\phi_0^2 \right)^2.
\label{lagr} \end{equation}
The field configuration describing a monopole is
\begin{equation}
\phi^\alpha=\phi_0 f(r) x^\alpha/r,
\label{ansatz} \end{equation}
while the metric can be written as
\begin{equation}
ds^2=-n^2(r)\, dt^2+b^2(r)\, dr^2+r^2(d\theta^2+\sin^2\theta\,d\chi^2+
\sin^2\theta\,\sin^2\chi\,d\phi^2).
\label{metric} \end{equation}
The Cartesian coordinates $x^\alpha$ are connected to 
the spherical coordinates $r$, $\theta$, $\chi$, $\phi$ through the 
standard relations. We also have $x^\alpha x_\alpha=r^2$ and 
$\phi^\alpha \phi_\alpha=\phi_0^2 f^2$. 

The Einstein equations are
\begin{eqnarray}
\frac{3}{b^2} \frac{1}{r}
\left( \frac{1}{r}-\frac{b'}{b} \right) &-& \frac{3}{r^2}
= \frac{1}{2M^3}T^0_{~0}
\nonumber \\
&=& \frac{1}{2M^3}\left[ 
\Lx-\frac{\phi_0^2 f'^2}{2b^2}-\frac{3\phi_0^2f^2}{2r^2}-U_B(f)
\right]
\label{ein00} \\
\frac{1}{b^2}\left[
\frac{1}{r}\left( \frac{1}{r}+2\frac{n'}{n}\right) 
-\frac{b'}{b}\left( \frac{n'}{n}+2\frac{1}{r}\right)
+\frac{n''}{n} \right]
&-&\frac{1}{r^2}=\frac{1}{2M^3}T^1_{~1}
\nonumber \\
&=&
\frac{1}{2M^3}\left[ 
\Lx-\frac{\phi_0^2 f'^2}{2b^2}-\frac{\phi_0^2f^2}{2r^2}-U_B(f)
\right]~~~~~~~~~
\label{einij} \\
\frac{3}{b^2}\frac{1}{r}\left(\frac{1}{r}+\frac{n'}{n} \right)
&-&\frac{3}{r^2} = \frac{1}{2M^3}T^4_{~4}
\nonumber \\
&=&
\frac{1}{2M^3}\left[ 
\Lx+\frac{\phi_0^2 f'^2}{2b^2}-\frac{3\phi_0^2f^2}{2r^2}-U_B(f)
\right], 
\label{ein44} 
\end{eqnarray} 
where primes indicate derivatives with respect to
$r$. Apart from the scalar field, we
have included a negative cosmological constant $-\Lx$ 
in the bulk.
An appopriate combination of the above equations gives the
equation of motion of the scalar field
\begin{eqnarray}
\frac{1}{b^2} f''&+&\frac{1}{b^2}\left[
\frac{3}{r}+\frac{(bn)'}{bn}-2\frac{b'}{b}
\right]f'-\frac{3}{r^2}f-\lx \phi_0^2(f^2-1)f
\nonumber \\
&=&\frac{1}{r^3}\left( \frac{r^3f'}{b^2} \right)'
+\frac{1}{b^2} \frac{(bn)'}{bn}f'
-\frac{3}{r^2}f-\lx \phi_0^2(f^2-1)f=0.
\label{eomf} \end{eqnarray}

We can rewrite eq. (\ref{ein00}) as 
\begin{eqnarray}
\frac{1}{b^2}&=&1+\frac{1}{12M^3}\Lx 
r^2-\frac{1}{6\pi^2M^3}\frac{\calm(r)}{r^2}
\label{b2} \\
\calm'&=&2\pi^2 r^3\left[ 
\frac{\phi_0^2 f'^2}{2b^2}+\frac{\phi_0^2f^2}{2r^2}+U_B(f)
\right].
\label{dmdr} 
\end{eqnarray}
By combining eqs. (\ref{ein00}), (\ref{ein44}) we obtain
\begin{equation}
\frac{(bn)'}{bn}=\frac{1}{6M^3}\phi_0^2rf'^2.
\label{bn} \end{equation}
The above equations cannot be solved analytically and one has to
integrate them numerically. However, 
the asymptotic form of the monopole configuration for $r\to \infty$ can
be deduced from eqs. (\ref{eomf})--(\ref{bn}). From eq. (\ref{eomf}) it
is apparent that $f\to 1$ for $r \to \infty$. 
It is clear then that the leading behaviour of $b^2(r)$ for large $r$ is
$b^{-2}(r)\to \Lx r^2/(12 M^3)$. Also, it can be seen
from eq. (\ref{bn}) that the term $\sim (bn)'$ in eq. (\ref{eomf}) is 
negligible for large $r$. In this way we find that $f\simeq 1+\beta/r^2$
for $r\to \infty$, with $\beta=-3/[2\lx\phi_0^2+\Lx/(3M^3)]$.

The asymptotic behaviour of the monopole solution implies that the
energy-momentum tensor takes the form
\begin{eqnarray}
{T}^{0}_{~0} = {T}^{4}_{~4} &=& \Lx -\frac{3\phi_0^2}{2r^2}
\label{t002b} \\
{T}^{1}_{~1} = {T}^{2}_{~2} = {T}^{3}_{~3} 
&=& \Lx -\frac{\phi_0^2}{2r^2}.
\label{t112b}
\end{eqnarray} 
Comparison with eqs. (\ref{t002a}), (\ref{t112a}) shows that we can
define an effective
integrated mass $\calm(r)=3\pi^2\phi_0^2r^2/2$. 
Contrary to our initial motivation of looking for an integrated 
mass $\sim r$, we have found
a stronger effect $\sim r^2$. 

In order to embed a brane, we consider a system of coordinates in
which the metric takes the form of eq. (\ref{sx2.1ex2}). 
The brane is located at $\eta=0$, while we
identify the half-space $\eta>0$ with the half-space
$\eta<0$. 
The time coordinate can be chosen such that $m(\tau,\eta=0)=1$.
Through an appropriate coordinate transformation 
\begin{equation}
t=t({\tau},\eta ),\qquad r=a({\tau},\eta )  \label{sx2.42exx}
\end{equation}
the metric (\ref{metric})
can be written in the form of eq. (\ref{sx2.1ex2}) with $k=1$. 
If we define $R(\tau)=a(\tau,\eta=0)$, we find that at the location 
of the brane
\begin{eqnarray}
\frac{\partial t}{\partial \tau} &=& 
\frac{1}{n(R)}\left[b^2(R)\dot{R}^2 +1\right]^{1/2}
\label{trans1} \\
\frac{\partial t}{\partial \eta} &=& 
-\frac{b(R)}{n(R)}\dot{R}
\label{trans2} \\
\frac{\partial a}{\partial \eta} &=& 
-\frac{1}{b(R)}\left[b^2(R)\dot{R}^2 +1\right]^{1/2},
\label{trans3} \end{eqnarray}
where the dot denotes a derivative with respect to $\tau$.

The presence of the brane induces an additional contribution 
to the energy-momentum tensor (as measured by an observer
comoving with the brane)
\begin{equation}
\left. T^A_{~C}\right|_b=\delta(\eta)~ {\rm diag} 
(-V-U_b(\phi)-\rht,-V-U_b(\phi)+\pt,
-V-U_b(\phi)+\pt,
-V-U_b(\phi)+\pt,0).
\label{deta} \end{equation}
The quantity $V$ is the brane tension, while the potential 
$U_b(\phi)$ accounts for possible interactions of the bulk field
with the brane. It is normalized so that $U_b(\phi_0)=0$.
The contributions $\rht$, $\pt$ arise from a perfect
fluid localized on the brane, with an equation of state 
$\pt=\pt(\rht)$.

The Einstein equations away from the brane remain unaffected,
while the contribution of eq. (\ref{deta}) 
can be taken into account by imposing appropriate boundary conditions
for the solutions of these equations.
In particular, the boundary conditions for the functions $m(\tau,\eta)$,
$a(\tau,\eta)$, $\phi(r(\tau,\eta))$ are 
\begin{eqnarray}
\left[ \frac{1}{m}\frac{\partial m}{\partial \eta}\right](\tau,\eta=0^+)&=&
\frac{1}{12M^3}(-V-U_b(\phi)+2\rht+3\pt)
\label{bound1} \\
\left[ \frac{1}{a}\frac{\partial a}{\partial \eta}\right](\tau,\eta=0^+)&=&
-\frac{1}{12M^3}(V+U_b(\phi)+\rht)
\label{bound2} \\
\frac{\partial \phi}{\partial \eta}(\tau,\eta=0^+)
=\frac{d\phi}{d r}\frac{\partial r}{\partial \eta}(\tau,\eta=0^+)
&=&\frac{1}{2}\frac{\partial U_b(\phi)}{\partial \phi}.
\label{bound3} \end{eqnarray}
For the monopole solution we derived above, eq. (\ref{bound3})
is satisfied for $\phi(R)\simeq \phi_0$ (or $f\simeq 1$) and large $R$, 
if the potential $U_b(\phi)$ has the form
\begin{equation}
U_b(\phi)=\left( \frac{\Lx}{3 M^3}\right)^{1/2} \left(\phi-\phi_0\right)^2.
\label{branepot} \end{equation}

The motion of the brane in an AdS bulk space with a global monopole is
expected to generate a mirage term in the Friedmann equation 
$\sim \calm(R)/R^4 \sim 1/R^2$: an effective curvature term.
The calculation of the brane cosmological evolution 
makes use of eqs. (\ref{bound1}), (\ref{bound2}) and proceeds in
complete analogy to refs. \cite{bulk1,bulk2}. 
As the bulk metric (\ref{metric})
has positive curvature along the three spatial dimensions parallel to
the brane, only a brane Universe with $k=1$ can be embedded in this background.
The values of the
bulk energy-momentum tensor at the location of the brane, as measured by
a brane observer, can be derived from eqs. (\ref{ein00})--(\ref{ein44})
and the transformations (\ref{trans1})--(\ref{trans3}).
For large values of the scale factor $R$ we find 
\begin{eqnarray}
H^2
&=&\left( \frac{\dot{R}}{R} \right)^2=
\frac{1}{6\mpl^2}\left[ \rht
+\beta^2 \phi^2_0 \left(\frac{\Lx}{3M^3}\right)^{1/2}  \frac{1}{R^4} \right]
-\frac{1}{R^2}+\frac{\phi_0^2}{4M^3}\frac{1}{R^2}
\label{hubble22a} \\
\dot{\rht}+3 H (\rht+\pt)&=&0.
\label{energ2a}
\end{eqnarray} 
The effective Friedmann equation (\ref{hubble22a}) is 
similar to that in
the conventional cosmology for a Universe with positive
spatial curvature. It includes, however, two additional contributions:
a) a term $\sim \phi^2_0 R^{-4}$ that arises from the brane 
potential $U_b(\phi)$, and b)
an effective curvature
term $\sim \phi_0^2 R^{-2}$, that arises through the influence of the 
bulk field (the monopole configuration) on the brane
evolution \footnote{Mirage curvature effects have also been 
found in ref. \cite{mirage}.}.
For large $R$ the last term 
can alter drastically the conventional picture: For
$\phi_0^2> 4M^3$ the Universe expands forever. 

We also point out that eq. (\ref{energ2a}), despite its simple form,
is the result of a subtle cancellation. The total energy density localized
on the brane includes the contribution of the potential $U_b(\phi)$, which
should appear in the l.h.s. of the conservation equation. 
However the r.h.s. of this equation is not zero, but equal to 
$-2 T^0_{~4}$, where $T^0_{~4}$ is the off-diagonal element of the 
bulk energy-momentum tensor at the location of the brane, 
as measured by a brane observer 
\cite{tet,bulk1,bulk2}. This can be computed to be 
\begin{equation}
T^0_{~4}=2\beta^2\phi_0^2\left( 
\frac{\Lx}{3M^3}\right)^{1/2} H \frac{1}{R^4}
\label{t05} \end{equation}
for this model. The two contributions cancel.
The physical interpretation is that the energy released as the field
$\phi(\tau,\eta=0)$ approaches $\phi_0$ is transferred to the 
bulk configuration.

\section{Summary and conclusions}

The purpose of this paper has been to demonstrate that the bulk
can induce various mirage effects on the brane, affecting significantly the
cosmological evolution even at low energy densities. 
We first summarized the well studied case of mirage radiation. 
We considered a generalized Vaidya-AdS metric in the bulk, which allows for
energy exchange between the bulk and the brane. 
We then pointed out that the same metric allows for a mirage component on
the brane with the equation of state of non-relativistic matter. 
The energy-momentum tensor of the anisotropic 
bulk fluid that can induce such an
effect has the form $T^0_{~0}=T^4_{~4}\sim -r^{-3}$, $T^1_{~1}=T^2_{~2}=
T^3_{~3}=0$. Because of the required negative pressure along
the extra dimension (${\rm p}=T^4_{~4}$) this fluid must correspond to
some field configuration. There is a strong
resemblance of the needed energy-momentum tensor
to that of a global monopole in four dimensions. The latter
has the form 
$T^t_{~t}\simeq T^r_{~r}\sim - r^{-2}$ and $T^\theta_{~\theta}
=T^\phi_{~\phi}\simeq 0$ 
at large distances $r$ from the center of the monopole. 

This motivated us to derive the cosmological evolution on a brane
wrapped around a global monopole in five dimensions. 
For large $r$ the resulting energy-momentum
tensor has the form $T^0_{~0}=T^4_{~4}\sim -3r^{-2}$, $T^1_{~1}=T^2_{~2}=
T^3_{~3}\sim -r^{-2}$. 
The pressure is negative and anisotropic.
The $r$-dependence arises because the leading contribution comes from the
angular part of the kinetic term in the action (the one that depends on
the angles parametrizing the three spatial directions parallel to the brane). 
On dimensional grounds this contribution is $\sim r^{-2}$, similarly to
what happens for the four-dimensional monopole. The integrated mass function
$\calm(r)$, defined in eq. (\ref{dmdr}), behaves as $\calm \sim r^2$ for
large $r$. It can be shown on
general grounds that the effective Friedmann equation on the brane receives a
contribution $\sim \calm(R)/R^4$, where $R$ is the scale factor. 
As a result, for a monopole-AdS bulk an
effective curvature term appears on the brane. Our explicit calculation
gives a term $\phi_0^2/(4 M^3 R^2)$, where $\phi_0$ is the asymptotic value of
the monopole field. For $\phi_0^2 > 4 M^3$, a brane Universe with positive
spatial curvature does not recollapse, but expands forever. 

Coming back to the question of the configuration in the bulk that can
induce a mirage term similar to cold dark matter, we point out that
the essential requirement is that the integrated  
mass function have the form $\calm(r)\sim r$. The energy-momentum tensor
$T^0_{~0}=T^4_{~4}\sim -r^{-3}$, $T^1_{~1}=T^2_{~2}=
T^3_{~3}=0$ that we discussed in section 2 satisfies this requirement. It
is also consistent with no energy exchange between the brane and the bulk.
As a result, the predicted brane evolution can be valid for arbitrarily
large $R$, as no brane energy is lost to the bulk. 
However, the physical interpretation of the required
bulk fluid is unclear.

In ref. \cite{bulk1} two other 
examples of mirage terms that scale $\sim R^{-3}$
were given. The bulk was assumed to be populated by a homogeneous
fluid with equation
of state $p=w\rho^\gamma$. For $w=1$, $\gamma=1/3$
or $\gamma=1$, $w=1/2$ the desired scale dependence was obtained. 
However, these models are consistent only with energy outflow from the 
brane. This implies that the brane energy density becomes zero at
a finite value of $R$, and the models lose their physical meaning 
for large $R$. 

In order to complete the picture, we also studied the evolution 
in the presence of a non-relativistic perfect fluid both on the brane
and in the bulk. In order to achieve this, we assumed that the
bulk can be described by the Tolman-Bondi-AdS metric. The bulk fluid 
is described by two arbitrary functions, that roughly correspond to the
initial distribution of matter and its velocity. The dynamical evolution
includes energy exchange between the brane and
the bulk, with outflow or inflow depending on the 
relative magnitude of the rate of expansion
on the brane (which can also be viewed as the motion of the brane within
the bulk) and the expansion rate of the bulk fluid.
The bulk matter induces a mirage component in the effective
Friedmann equation for the brane expansion, 
with positive energy density $\rho_{eff}$
and pressure $p_{eff}>\rho_{eff}/3$. This energy density 
falls off faster than $R^{-4}$.

\vspace {0.5cm}
\noindent{\bf Acknowledgments}\\
\noindent One of the authors (E.N.S.) acknowledges the financial support of the National Scholarship Foundation (I.K.Y.). 
This work was partially supported through the RTN contract MRTN--CT--2004--503369
of the European Union, and the research programs ``Pythagoras'' of the 
Ministry of National Education grant no 70-03-7310 and ``Kapodistrias'' of the University of Athens.


\vskip 1.5cm

\begin{thebibliography}{999}

\bibitem{rs} 
L. Randall and R. Sundrum, 
Phys. Rev. Lett. {\bf 83} (1999) 3370 
[arXiv:hep-th/9905221];
Phys. Rev. Lett. {\bf 83} (1999) 4690 
[arXiv:hep-th/9906064]. 
\bibitem{binetruy}
P.~Binetruy, C.~Deffayet and D.~Langlois,
Nucl.\ Phys.\ B {\bf 565} (2000) 269
[arXiv:hep-th/9905012];
\\
P.~Binetruy, C.~Deffayet, U.~Ellwanger and D.~Langlois,
Phys.\ Lett.\ B {\bf 477} (2000) 285
[arXiv:hep-th/9910219].

\bibitem{csaki}
C.~Csaki, M.~Graesser, C.~F.~Kolda and J.~Terning,
Phys.\ Lett.\ B {\bf 462} (1999) 34
[arXiv:hep-ph/9906513];
\\
J.~M.~Cline, C.~Grojean and G.~Servant,
Phys.\ Rev.\ Lett.\  {\bf 83} (1999) 4245
[arXiv:hep-ph/9906523]; 
\\
A. Padilla, \emph{Braneworld Cosmology and Holography}, Ph.D. thesis, (2002) University of Durham, England 
[arXiv:hep-th/0210217].

\bibitem{kraus}
P.~Kraus,
JHEP {\bf 9912} (1999) 011
[arXiv:hep-th/9910149].
\bibitem{brax}
P.~Brax and C.~van de Bruck,
Class.\ Quant.\ Grav.\  {\bf 20} (2003) R201
[arXiv:hep-th/0303095];
\\
R.~Maartens, \emph{Living Rev. Rel.} \textbf{7}, 7 
(2004) 
[arXiv:gr-qc/0312059];
\\
E.~Kiritsis,
Fortsch.\ Phys.\  {\bf 52} (2004) 200
[arXiv:hep-th/0310001].
\bibitem{mirage}
A.~Kehagias and E.~Kiritsis,
JHEP {\bf 9911} (1999) 022
[arXiv:hep-th/9910174].
 
\bibitem{hebecker} 
A.~Hebecker and J.~March-Russell,
Nucl.\ Phys.\ B {\bf 608} (2001) 375
[arXiv:hep-ph/0103214].

\bibitem{supergravity}
  A.~Lukas, B.~A.~Ovrut, K.~S.~Stelle and D.~Waldram,
  Nucl.\ Phys.\ B {\bf 552} (1999) 246
  [arXiv:hep-th/9806051];\\
  J.~R.~Ellis, Z.~Lalak, S.~Pokorski and S.~Thomas,
  Nucl.\ Phys.\ B {\bf 563} (1999) 107
  [arXiv:hep-th/9906148];\\
  M.~Gunaydin and M.~Zagermann,
  Nucl.\ Phys.\ B {\bf 572} (2000) 131
  [arXiv:hep-th/9912027]; 
  Phys.\ Rev.\ D {\bf 62} (2000) 044028
  [arXiv:hep-th/0002228];\\
  A.~Ceresole and G.~Dall'Agata,
  Nucl.\ Phys.\ B {\bf 585} (2000) 143
  [arXiv:hep-th/0004111];\\
  A.~Falkowski, Z.~Lalak and S.~Pokorski,
  Phys.\ Lett.\ B {\bf 509} (2001) 337
  [arXiv:hep-th/0009167]; \\
  E.~Bergshoeff, R.~Kallosh and A.~Van Proeyen,
  JHEP {\bf 0010} (2000) 033
  [arXiv:hep-th/0007044]; \\
  G.~A.~Diamandis, B.~C.~Georgalas, P.~Kouroumalou and A.~B.~Lahanas,
  Phys.\ Lett.\ B {\bf 602} (2004) 112
  [arXiv:hep-th/0402228].


\bibitem{vandebruck}
C.~van de Bruck, M.~Dorca, C.~J.~A.~Martins and M.~Parry,
Phys.\ Lett.\ B {\bf 495} (2000) 183
[arXiv:hep-th/0009056].
\bibitem{induced}
H.~Collins and B.~Holdom,
Phys.\ Rev.\ D {\bf 62} (2000) 105009
[arXiv:hep-ph/0003173];
\\
E.~Kiritsis, N.~Tetradis and T.~N.~Tomaras,
JHEP {\bf 0203} (2002) 019
[arXiv:hep-th/0202037];
\\
K.~i.~Maeda, S.~Mizuno and T.~Torii,
Phys.\ Rev.\ D {\bf 68} (2003) 024033
[arXiv:gr-qc/0303039].
\bibitem{gauss}
N.~Deruelle and T.~Dolezel,
Phys.\ Rev.\ D {\bf 62} (2000) 103502
[arXiv:gr-qc/0004021];
\\
B.~Abdesselam and N.~Mohammedi,
Phys.\ Rev.\ D {\bf 65} (2002) 084018
[arXiv:hep-th/0110143];
\\
C.~Germani and C.~F.~Sopuerta,
Phys.\ Rev.\ Lett.\  {\bf 88} (2002) 231101
[arXiv:hep-th/0202060];
\\
C.~Charmousis and J.~F.~Dufaux,
Class.\ Quant.\ Grav.\  {\bf 19} (2002) 4671
[arXiv:hep-th/0202107];
\\
G.~Kofinas, R.~Maartens and E.~Papantonopoulos,
JHEP {\bf 0310} (2003) 066
[arXiv:hep-th/0307138];

S. Nojiri and S. D. Odintsov, JHEP \textbf{0007} (2000) 049 [arXiv:hep-th/0006232]; 

J. P. Gregory and A. Padilla, Class. Quant. Grav. \textbf{20} (2003) 4221 [arXiv:hep-th/0304250].

\bibitem{tet}
E.~Kiritsis, G.~Kofinas, N.~Tetradis, T.~N.~Tomaras and V.~Zarikas,
JHEP {\bf 0302} (2003) 035
[arXiv:hep-th/0207060];
\\
N.~Tetradis,
Phys.\ Lett.\ B {\bf 569} (2003) 1
[arXiv:hep-th/0211200].
\bibitem{tomaras}
T.~N.~Tomaras,
arXiv:hep-th/0404142;\\ E.~Kiritsis, Fortsch.\ Phys.\  {\bf 52} (2004) 568.
\bibitem{bulk1}
P.~S.~Apostolopoulos and N.~Tetradis,
Class.\ Quant.\ Grav.\  {\bf 21} (2004) 4781
[arXiv:hep-th/0404105].
\bibitem{bulk2}
N.~Tetradis,
Class.\ Quant.\ Grav.\  {\bf 21} (2004) 5221
[arXiv:hep-th/0406183].

\bibitem{bulk3}
V. P. Frolov, M. Snajdr and D. Stojkovic, Phys. Rev. D \textbf{68} (2003) 044002 [arXiv:gr-qc/0304083];
J. P. Gregory and A. Padilla, Class. Quant. Grav. \textbf{19} (2002) 4071 [arXiv:hep-th/0204218]; 
A. N. Aliev and A. E. G\"umr\"uk\c {c}\"uo\~glu, Class. Quant. Grav. \textbf{21} (2004) 5081 [arXiv:hep-th/0407095]

\bibitem{birm}
D.~Birmingham,
Class.\ Quant.\ Grav.\  {\bf 16} (1999) 1197
[arXiv:hep-th/9808032].
\bibitem{gregory}
S.~Mukohyama, T.~Shiromizu and K.~i.~Maeda,
Phys.\ Rev.\ D {\bf 62} (2000) 024028
[Erratum-ibid.\ D {\bf 63} (2001) 029901]
[arXiv:hep-th/9912287];
\\
P.~Bowcock, C.~Charmousis and R.~Gregory,
Class.\ Quant.\ Grav.\  {\bf 17} (2000) 4745
[arXiv:hep-th/0007177].


\bibitem{wang}
P.~C.~Vaidya,
Proc.\ Ind. \ Acad. \ Sci. A {\bf 33} (1951) 264;
{\it reprinted}
Gen.\ Relativ.\ Grav. {\bf 31} (1999) 121;
\\
A.~z.~Wang and Y.~m.~Wu,
Gen.\ Relativ.\ Grav. {\bf 31} (1999) 107.


\bibitem{vaidya}
D.~Langlois, L.~Sorbo and M.~Rodriguez-Martinez,
Phys.\ Rev.\ Lett.\  {\bf 89} (2002) 171301
[arXiv:hep-th/0206146];
\\
E.~Leeper, R.~Maartens and C.~F.~Sopuerta,
Class.\ Quant.\ Grav.\  {\bf 21} (2004) 1125
[arXiv:gr-qc/0309080].



\bibitem{general}
P.~S.~Apostolopoulos and N.~Tetradis,
Phys.\ Rev.\ D {\bf 71} (2005) 043506 [arXiv:hep-th/0412246].


\bibitem{vilenkin}
M.~Barriola and A.~Vilenkin,
Phys.\ Rev.\ Lett.\  {\bf 63} (1989) 341.


\bibitem{tb}
R.~C.~Tolman,
Proc.\ Nat.\ Acad.\ Sci.\  {\bf 20} (1934) 169;
\\
H.~Bondi,
Mon.\ Not.\ Roy.\ Astron.\ Soc.\  {\bf 107} (1947) 410.

\bibitem{brito}
F.~A.~Brito, F.~F.~Cruz and J.~F.~N.~Oliveira,
Phys.\ Rev.\ D {\bf 71} (2005) 083516 [arXiv:hep-th/0502057].

\end{thebibliography}
\end{document}